Determination of the vibrational contribution to the entropy change at the martensitic transformation in Ni-Mn-Sn metamagnetic shape memory alloys: A combined approach of time-of-flight neutron spectroscopy and ab-initio calculations

V. Recarte,[1,2] M. Zbiri,[3] M. Jiménez-Ruiz,[3] V. Sánchez-Alarcos[1,2] and J.I. Pérez-Landazábal,[1,2]

[1]*Departamento de Física, Universidad Pública de Navarra, Campus de Arrosadía, 31006 Pamplona, Spain*
[2]*Institute for Advanced Materials (INAMAT), Universidad Pública de Navarra, Campus de Arrosadía, 31006 Pamplona, Spain*
[3]*Institut Laue-Langevin (ILL), 71 Avenue des Martyrs, CS 20156, 38042 Grenoble Cedex 9, France*

**Abstract**

The different contributions to the entropy change linked to the austenite-martensitic transition in a Ni-Mn-Sn metamagnetic shape memory alloy have been determined by combining different experimental techniques. The vibrational contribution has been inferred from the vibrational density of states of both the martensitic and austenite phases. This has been accomplished by combining time-of-flight neutron scattering measurements and *ab-initio* calculations. Further, the electronic part of the entropy change has also been calculated. Since the martensitic transformation takes place between two paramagnetic phases, the magnetic contribution can be neglected and the entropy change can be reduced to the sum of two terms: vibrational and electronic. The obtained value of the vibrational contribution ($-36 \pm 5 \text{ J kg}^{-1}\text{K}^{-1}$) nearly provides the total entropy change measured by calorimetry ($-41 \pm 3 \text{ J kg}^{-1}\text{K}^{-1}$), the difference being the electronic contribution within the experimental error.

**1. Introduction**

Ni-Mn-based Heusler ferromagnetic shape memory alloys (FSMAs) have attracted much attention due to their potential application as multiferroic materials. Their functional properties are linked to the presence of the martensitic transformation (MT), a diffusionless first order phase transition from a high temperature cubic structure (austenite) to a low temperature structure (martensite), showing a strong coupling between magnetic and structural variables [1]. Unlike Ni-Mn-Ga alloys, the so-called metamagnetic systems, Ni-Mn-Z (Z=In, Sn, Sb), show a drastic change in magnetization at the MT, being significantly smaller in the weak magnetic martensite phase than in the ferromagnetic austenite phase [2]. This feature makes the MT very sensitive to external magnetic fields. Metamagnetic shape memory effect [3], inverse magnetocaloric effect [4], and giant magnetorresistance effect [5] have been reported in these systems.

Structural, electronic and magnetic effects are believed to play a predominant role to trigger the martensitic transformation. For example, the dynamical instability of the austenite phase prior to the MT commonly gives rise to precursor effects such as the softening of the low-lying transverse $TA_2$-phonon branch [6-10], or the anomalous behavior of the *C'* shear modulus [11-13]. Concerning the origin of the MT it was ascribed to a Jahn-Teller effect [14], since a redistribution of the electronic density of states (DOS) was observed upon the martensitic transformation in $Ni_2MnGa$. In this context, a reduction of the electronic DOS close to the Fermi energy, when transforming from the austenite to the martensite, was predicted by band structure calculations in Ni-Mn-Ga [15-17] and Ni-Mn-Sn [18], and confirmed by photoemission spectroscopy [17-19] and EXAFS measurements [20]. Empiric linear dependences between the MT temperature and the concentration of valence electron per unit cell *e/a* were found for several Heusler based FSMAs, which corroborate the role of the electronic subsystem on the MT in these compounds [21-23].

Magnetism plays a key role in the stability of the austenite phase towards further transitions, demonstrating that a strong magneto-elastic interaction is necessary for the induction of the intermediate phase in $Ni_{50}Mn_{25}Ga_{25}$ [24-26]. Additionally, the increase of the atomic order degree stabilizes the structural phase exhibiting a higher magnetic moment, as a result of the effect of the magnetic exchange coupling variations on the free energy difference between the austenite and martensite phases [27].

Therefore, any discussion about the characteristics of the MT through the entropy change must consider the sum of the changes of the three components; vibrational, magnetic and electronic: $\Delta S_{MT} = \Delta S_{vib} + \Delta S_{mag} + \Delta S_{el}$. Usually the dominant role has been attributed to the vibrational and magnetic contributions, disregarding the electronic entropy change [28-31]. Nevertheless, the proper evaluation of the different contributions to MT entropy change remains an open question that needs to be addressed.

It is the purpose of this work to analyze and accurately quantify the vibrational and electronic contributions to the MT in a Ni-Mn-Sn metamagnetic shape memory alloy. To reach this goal we used a combination of time-of-flight neutron scattering measurements and ab-initio calculations. The calculations have been carried out for the sake of the analysis of the neutron-determined generalized density of states, of both the austenite and martensite phases. A generalized density of states (GDOS) is the phonon spectrum measured from inelastic neutron scattering. In contrast to the vibrational density of states, relevant to the present work on entropy, the GDOS involves a weighting of the scatterers (ions) with their scattering powers [32]. For a polyatomic system, ab-initio calculations are a robust tool which allowed us to infer accurately the vibrational contribution from the (generalized) neutron-weighted counterpart. The comparison of the total entropy change measured by calorimetry, and the electronic and magnetic entropy changes determined from their respective densities of states, helped in performing a detailed and relevant analysis.

## 2. Experimental details

A polycrystalline alloy of $Ni_{50}Mn_{36}Sn_{14}$ was prepared from high purity (99.98%) elements by arc melting under protective Ar atmosphere. The ingot was re-melted six times and homogenized in vacuum quartz ampoules during 24 hours at a high enough temperature (1173 K) to promote the diffusion but avoiding the loss of manganese. Finally, the ingot was annealed two hours at 1273 K followed by quenching into ice water in order to avoid decomposition and to retain the austenite phase. The composition was checked by energy-dispersive spectroscopy in a JEOL JSM-5610LV scanning electron microscope. Neutron diffraction measurements at 400 K, not shown, reveal the presence of a single phase, the cubic L2$_1$ austenite. The structural and magnetic transformations were determined from low field magnetization measurements performed in a QD MPMS XL-7 SQUID magnetometer and DSC measurements in a TA Q100 calorimeter. Inelastic neutron scattering measurements on ~ 3 grams of a polycrystalline sample of Ni-Mn-Sn were performed on the direct-geometry, cold-neutron time-of-flight time-focusing spectrometer IN6, at the Institut Laue Langevin (Grenoble, France). Data were collected at 250 and 400 K, in the up-scattering mode (neutron energy gain), using an incident neutron wavelength $\lambda_i$=5.12 Å, leading to a resolution of 0.07 meV at the elastic line. Data were analyzed after applying standard corrections; including detector efficiency calibration, background subtraction, and energy-dependent detector efficiency correction.

## 3. Results and discussion

The forward martensitic transformation (FMT) start and finish temperatures, $M_s$=335 K and $M_f$=290 K, respectively, and the reverse martensitic transformation (RMT) start and finish temperatures, $A_s$=305 K and $A_f$=350 K, respectively, were determined by DSC measurements, as shown in Figure 1-a. The entropy change linked to the MT can be calculated as $\Delta S_{MT} = \int_{T_s}^{T_f}(1/T)(dQ/dt)\dot{T}^{-1}dT$, where $dQ/dt$ is the heat flow interchanged by the sample per unit mass (W/g in Figure 1-a), $\dot{T}$ is the heating/cooling rate (0.166 K s$^{-1}$), and $T_s$ and $T_f$ are the transformation start and finish temperatures, respectively. After removing the baseline and integrating the peaks shown in Figure 1-a, the average between the forward and reverse transitions gives a value of $\Delta S_{MT} = -41 \pm 3$ J kg$^{-1}$K$^{-1}$, for the martensitic transition. The sequence of magnetostructural transformations has been determined from the low-field magnetization measurements (H=100 Oe), shown in Figure 1-b. The magnetization increases on cooling below the austenite Curie temperature, $T_C^{aust}$, and drastically decreases on further cooling below the forward MT temperature. The structural and magnetic transitions occur so close to each other that the magnetization increase associated with the ferromagnetic ordering is truncated by the sudden appearance of the paramagnetic martensite, which orders magnetically below the martensitic Curie temperature, $T_C^{mart}$. From the inverse of the magnetic susceptibility (bottom inset in Figure 1-b), the magnetic ordering temperatures of the austenite and martensitic phases have been estimated to be close to 310 K and 240 K, respectively. In order to confirm these values we carried out modulated DSC measurements (top inset in Figure 1-b) from which values of $T_C^A$=300 K and $T_C^M$=225 K have been determined. According to the temperature sequence, $M_s$=335 K, $T_C^A$=300 K, $M_f$=290 K and $T_C^M$=225 K, we consider that the assumption of a MT between paramagnetic phases could be valid since only the last fraction of transforming austenite is ferromagnetic. On the basis of these observations, the magnetic contribution to the entropy change can be disregarded. Therefore, only the vibrational and electronic components contribute mainly to the total entropy change, which can be reduced to $\Delta S_{MT} = \Delta S_{vib} + \Delta S_{el}$.

The *Q*-averaged, one-phonon generalized phonon density of states (GDOS) was obtained using the incoherent approximation in the same way as in previous works dealing with phonon dynamics [33,34]. In the incoherent one-phonon approximation, the measured scattering function *S(Q,E)*, as observed in inelastic neutron experiments, is related [35,36] to the phonon generalized density of states $g^{nw}(E)$, as seen by neutrons, as follows:

$$g^{nw}(E) = A < \frac{e^{2W_i(Q)}}{Q^2} \frac{E}{n_T(E)+\frac{1}{2}\pm\frac{1}{2}} S(Q,E) > \qquad (1)$$

With:

$$g^{nw}(E) = B \sum_i \left\{\frac{4\pi b_i^2}{m_i}\right\} x_i g_i(E) \qquad (2)$$

where the + or − signs correspond to energy loss or gain of the neutrons respectively and $n_T(E)$ is the Bose-Einstein distribution. *A* and *B* are normalization constants and $b_i$, $m_i$, $x_i$, and $g_i(E)$ are, respectively, the neutron scattering length, mass, atomic fraction, and partial density of states of the i[th] atom in the unit cell. The quantity between < > represents suitable average over all *Q* values at a given energy. $2W(Q)$ is the Debye-Waller factor. The weighting factors $\frac{4\pi b_i^2}{m_i}$ for various atoms in the units of barns/amu are [37]: Ni: 0.315; Mn: 0.039 and Sn: 0.041.

Figure 2 shows the measured "neutron weighted" phonon DOS, $g^{nw}(E)$, of the sample at 400K (the paramagnetic austenite phase) and at 250 K (the paramagnetic martensite phase). A faster Debye growth was observed in the martensite phase as compared to the austenite case, which points towards the presence of a residual paramagnetic scattering in the martensite phase [33]. In order to correct the effect of the deviation from a quadratic Debye-like behavior in the martensite case, we have adjusted the low-energy part to follow a quadratic law [33]. The main difference between the phonon DOS of both structures is the peak around 10 meV, characterizing the austenite phase. This peak is related to the presence of soft modes along the low-lying transverse TA$_2$-phonon branch, which is common to the cubic phase in the shape memory alloys, based on Heusler compounds [1]. The martensitic phase shows a slight evidence of this peak which could indicate a possible existence of soft modes in this phase. The true (vibrational) total phonon DOS of the alloy can be written following Equation 2, but without considering the neutron-dependent ionic scattering powers:

$$g(E) = \sum_i x_i\, g_i(E) \qquad (3)$$

The function $g(E)$ is required for calculating thermodynamic quantities, such as the phonon entropy or phonon contribution to heat capacity. The vibrational entropy of each phase $S_{vib}(T)$ at a temperature $T$ is given in the quasiharmonic approximation by [38]:

$$S_{vib}(T) = 3k_B \int_0^{E_{max}} [(n_T(E) + 1)\ln(n_T(E) + 1) - n_T(E)\ln(n_T(E))]\, g(E)\, dE \qquad (4)$$

where $E_{max}$ is the phonon energy cutoff.

The next step is to infer $g(E)$ from $g^{nw}(E)$. However this is straightforward only when the values of $4\pi b_i^2/m_i$ are similar for all the atoms, so $g^{nw}(E) \approx g(E)$ and no further corrections are necessary. But presently, as above indicated, the ionic neutron weighting factors are different: Ni: 0.315; Mn: 0.039 and Sn: 0.041, the Ni contribution being one order of magnitude higher. So the measured "neutron weighted" DOS can be corrected to obtain the true (vibrational) DOS. Therefore, *ab-initio* calculations represent a viable and accurate route to estimate a correction function $f(E)$, in such a way that $g(E) = f(E)\, g^{nw}(E)$.

Relaxed geometries and total energies were obtained using the projector-augmented wave (PAW) formalism [39,40] of the spin-polarized Kohn-Sham density functional theory [41,42], within both the local density approximation (LDA) and the generalized gradient approximation (GGA), implemented in the Vienna *ab-initio* simulation package (VASP) [43]. The GGA was formulated by the Perdew–Burke–Ernzerhof (PBE) density functional [44]. The LDA was based on the Ceperly–Alder parametrization by Perdew and Zunger [45]. The valence electronic configurations of Ni, Mn and Sn as used for pseudo potential generation are $d^8s^2$, $d^6s^1$ and $s^2p^2$, respectively. All results were well converged with respect to *k*-mesh and energy cutoff for the plane wave expansion. A plane wave energy cutoff of 400 eV was used, and the integrations over the Brillouin zone were sampled on a 3×3×3 grid of k-points generated by Monkhorst-pack method [46], for the supercell phonon calculations. The break conditions for the self-consistent field (SCF) and for the ionic relaxation loops were set to $10^{-8}$ eV and $10^{-5}$ eV Å$^{-1}$, respectively. The latter break condition means that the obtained Hellmann–Feynman forces are less than $10^{-5}$ eV Å$^{-1}$.

Calculations were carried out on the stoichiometric material Ni$_2$MnSn; under both the high-temperature austenite (cubic) and low-temperature martensitic (orthorhombic) phases; with 4 formula units (16 atoms) per unit cell in each phase. The cubic structure (space group $Fm\bar{3}m$) [47], contains 3 crystallographically inequivalent atoms (1 Ni, 1 Mn, and 1 Sn), whereas the orthorhombic phase (space group Pmma) [48], contains 6 crystallographically inequivalent atoms (2 Ni, 2 Mn, and 2 Sn). Table 1 gathers the related crystallographic data used as starting structures in the calculations.

In order to determine all force constants, the supercell approach was used for lattice dynamics calculations. Thus, 2x2x2 and 1x2x2 supercells were constructed from the relaxed austenite and martensite structures, respectively. In the former phase the supercell contains 32 formula units (128 atoms), whereas in the latter phase the supercell contains 16 formula units (64 atoms). Total energies and Hellmann-Feynman forces were calculated for 6 and 36 structures resulting from individual displacements of the symmetry inequivalent atoms in the austenite (cubic) and martensite (orthorhombic) supercells, respectively, along the six inequivalent Cartesian directions (±*x*, ±*y* and ±*z*). Phonon modes were extracted from subsequent calculations using the direct method [49] as implemented in the PHONON software [50].

Figure 3 shows typical calculated vibrational densities of states (vDOS) of the austenite phase of Ni$_2$MnSn, using both GGA (PBE) and LDA, from the direct method whose numerical procedure is described above. The perfect match of our PBE-estimated vDOS with the PBE-predicted one in the literature, determined from DFPT calculations [51], is worth noticing. Further, our LDA-based vDOS illustrates the expected volume effect of the exchange-correlation in terms of the maximum energy transfer which in the case of LDA exceeds that from PBE calculations (~ 32 meV from LDA and ~ 28 meV from our PBE calculations and reference [51]).

The correction function, $f(E)$, has been extracted from *ab-initio* calculations as follows: after calculating the partial density of states for both structures, the vibrational $g_{calc}(E)$ and generalized $g_{calc}^{nw}(E)$ DOS of the alloy have been estimated, according to equations (3) and (2), respectively. We considered the alloy composition, i.e. $x_{Ni} = 0.50, x_{Mn} = 0.36$ and $x_{Sn} = 0.14$. Then, $f(E)$ has been inferred as the following ratio: $f(E) = g_{calc}(E)/g_{calc}^{nw}(E)$. Finally $g(E)$ has been determined by multiplying the measured (generalized) neutron weighted $g^{nw}(E)$ by $f(E)$.

Figures 4–a and 4-b show the vibrational DOS of the austenite and martensite, at 400 K and 250 K respectively, after applying the *ab-initio* determined correction function *f(E)*, using PBE and LDA methods, respectively. According to the displacive character of the MT, the corrections are very similar for both phases. In addition, both methods give coherent results in such a way that the obtained phonon DOS show the same shape and features. Having determined the vibrational density of states, we can calculate the vibrational contribution to the entropy change at the MT: $\Delta S_{vib}(T_{MT}) = S_{vib}^{mart}(T_{MT}) - S_{vib}^{aust}(T_{MT}) = -36 \pm 5\, J\, \text{kg}^{-1}\text{K}^{-1}$. The entropy of each phase has been calculated using Equation (4) at T=310 K, which is the temperature of the DSC peak for the FMT. This value is the average of the values obtained by LDA and PBE calculations. The negative contribution is in agreement with the reduction in the vibrational entropy due the transformation of an instable open structure to a close-packed one. In order to check the significance of the correction, the neutron-weighted entropy change has also been calculated by considering in Equation (4) the generalized neutron-weighted DOS (Figure 2). A value of $\Delta S_{vib}^{nw}(T_{MT}) = -40 \text{ J kg}^{-1} \text{ K}^{-1}$ is obtained, and which is found to be very close to the neutron-weighting factor corrected one. This points towards a minor effect of the neutron scattering power of the atomic components of this shape memory alloy on the vibrational contribution to $\Delta S$, at the MT, reflecting the weak character of the applied correction although is not negligible in the high energy part of the phonon spectra. As the integral (Equation 4) is modulated by the Bose-Einstein distribution, the low energy modes contribute the most to the vibrational entropy. The small difference between the total entropy change $\Delta S_{MT} = -41 \pm 3 \text{ J kg}^{-1} \text{K}^{-1}$ and the vibrational contribution $\Delta S_{vib} = -36 \pm 5 \text{ J kg}^{-1} \text{K}^{-1}$ could be related to the electronic contribution. We can estimate the electronic entropy change at the MT as $\Delta S_{el}(T_{MT}) = S_{el}^{mart}(T_{MT}) - S_{el}^{aust}(T_{MT})$. In a recent work, the electronic DOS of the austenite and a tetragonal martensite have been calculated in a Ni$_{50}$Mn$_{37.5}$Sn$_{12.5}$ alloy [18]. DFT calculations show the presence of a peak in the electronic DOS in the vicinity of $E_F$ for the austenite, which does not appear in the martensite phase. This excess of electronic states destabilizes the cubic structure. In spite of the fact that the true structure of the martensite is orthorhombic, the values of the electronic DOS at the Fermi energy, $n(E_F)$, calculated in [18] can be used to estimate the electronic contribution by applying the Sommerfeld expansion, leading to a value of $\Delta S_{el}(T_{MT}) \approx -2 \text{ J kg}^{-1} \text{K}^{-1}$. Hence, the small difference between $\Delta S_{MT}$ and $\Delta S_{vib}$ can be ascribed to an electronic origin. Although the instability of the cubic phase has an electronic origin, our findings indicate that the main internal energy reduction at the MT happens in the lattice subsystem. The electronic contribution appears to be a secondary contribution to the transformation entropy change.

**4. Conclusions**

The vibrational densities of states (DOS) in the austenitic and martensitic phases of a Ni-Mn-Sn metamagnetic shape memory alloy have been determined by means of time-of-flight inelastic neutron scattering measurements and *ab-initio* calculations. From the obtained DOS, the vibrational contribution to the entropy change at the martensitic transformation has been calculated ($\Delta S_{vib} = -36$ J kg$^{-1}$K$^{-1}$). It represents approximately 90% of the total entropy change at the transformation. Taking into account that the martensitic transformation takes place between two paramagnetic phases (and therefore no magnetic contribution is expected), the electronic contribution is inferred to be about 10% of the total entropy change. Our results highlight the predominant role of the vibrational entropy in driving the martensitic transformation.

**Acknowledgments**

This work has been carried out with the financial support of the Spanish "Ministerio de Economía y Competitividad" (Project number MAT2012-37923-C02-01) and FEDER funds. The Institut Laue-Langevin (ILL) facility, Grenoble, France, is acknowledged for providing beam time on the IN6 spectrometer.

**Table Captions:**

Table 1: Crystallographic data of the austenite and martensitic phases of a stoichiometric model system, Ni$_2$MnSn, used as starting structures in the ab-intio lattice dynamical calculations.

| Phase | Site | Atom |
|---|---|---|
| Austenite (Cubic, $Fm\bar{3}m$) [46] a=b=c=5.9690 Å | *4a* (0 0 0) | Mn |
| | *4b* (1/2 1/2 1/2) | Sn |
| | *8c* (1/4 1/4 1/4) | Ni |
| Martensitic (Orthorhombic, Pmma) [47] a=8.5837 Å b=5.6021 Å c=4.3621 Å | *2a* (0 0 0) | Mn |
| | *2f* (1/4 1/2 0.574) | Mn |
| | *2b* (0 1/2 0) | Sn |
| | *2e* (1/4 0 0.562) | Sn |
| | *4h* (0 0.2495 1/2) | Ni |
| | *4k* (1/4 0.2485 0.0913) | Ni |

Table 1

**Figure captions:**

Figure 1: a) DSC cooling-heating cycle in the vicinity of the martensitic tranformation. b) Low-field magnetization cooling-heating cycle between 10 K and 375 K. The top inset, in this figure, shows the magnetic ordering temperatures determined by modulated DSC, whereas the bottom inset shows the inverse of the magnetic susceptibility.

Figure 2: Neutron-weighted (generalized) phonon density of states of the austenite and martensite phases of Ni-Mn-Sn, measured at 400 K and 250K, respectively.

Figure 3: Vibrational density of states (DOS) from LDA (a) and PBE (b) calculations of the austenite phase of the stoichiometric $Ni_2MnSn$. The effect of the exchange-correlation can clearly be seen when comparing LDA (a) and GGA (PBE) (b) results. In the former case the maximum energy transfer extends to a higher energy value (7.6 THz ~ 32 meV) than that in the latter case (6.8 THz ~ 28 meV).

Figure 4: Vibrational densities of states for austenite and martensite, at 400 K and 250 K respectively, inferred from the measured spectra, after applying the ab-initio calculated neutron weighting correction function *f(E)* using PBE (a) and LDA (b) methods.

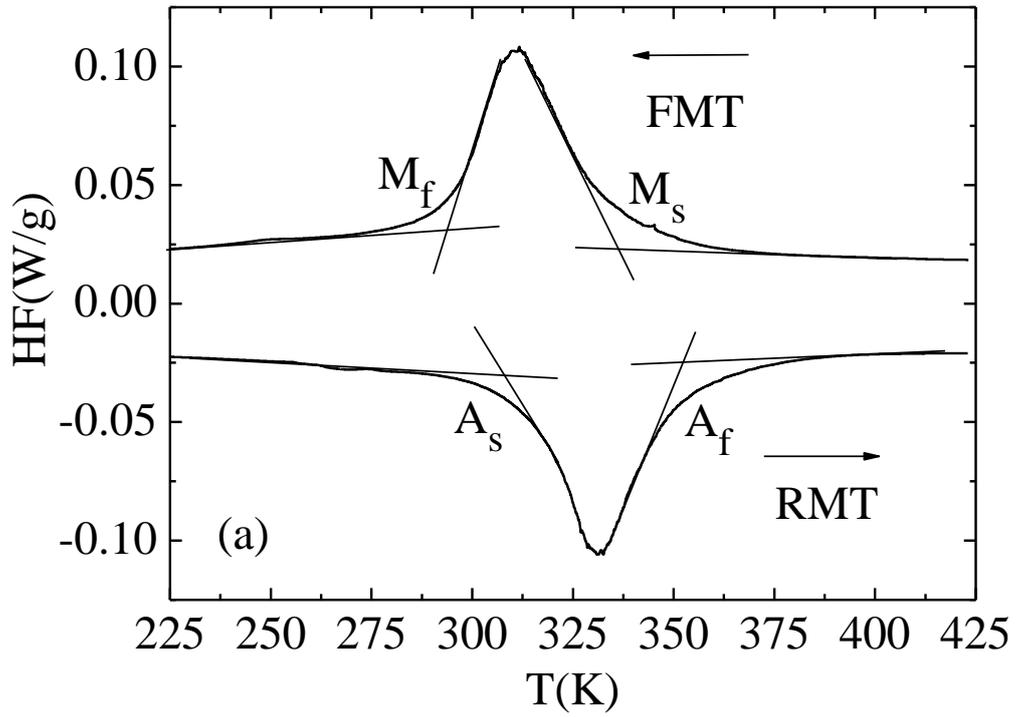

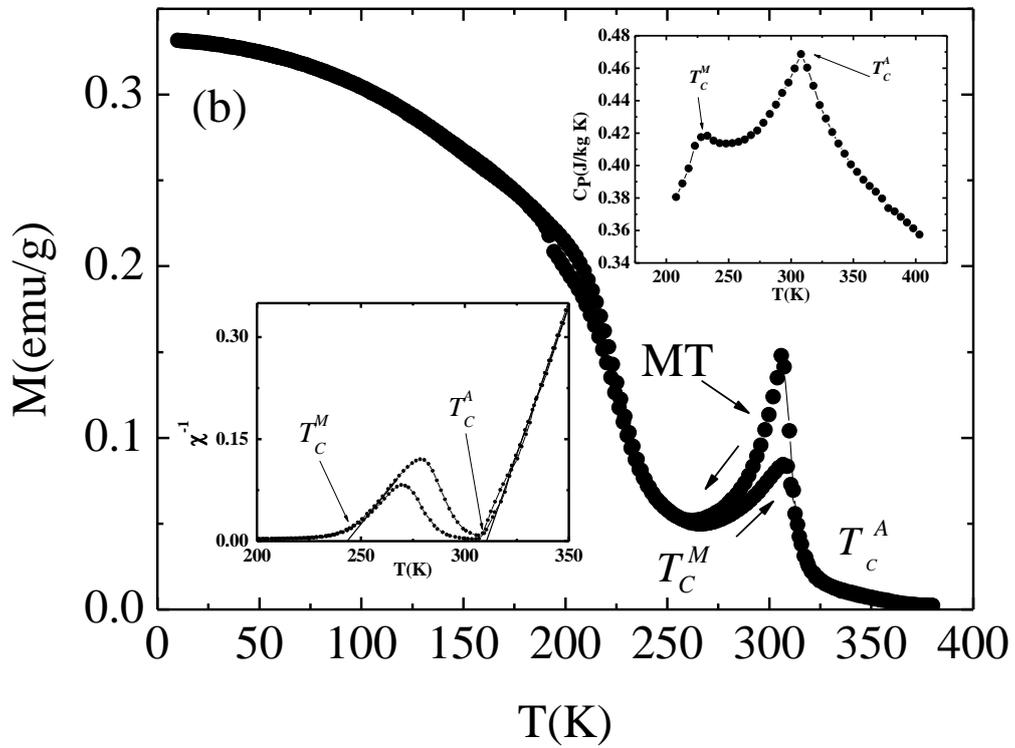

Figure 1

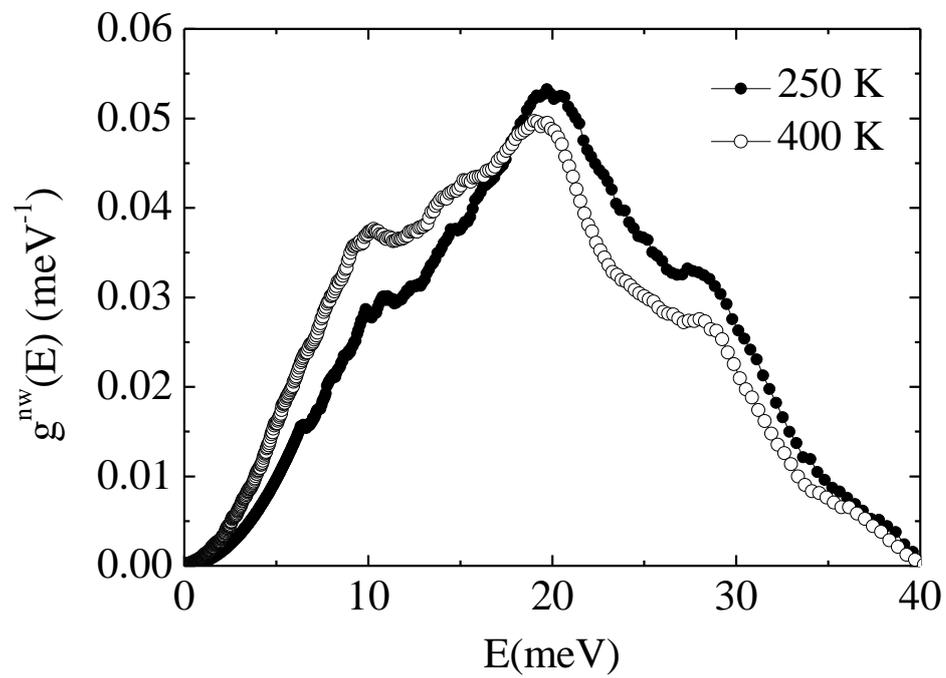

Figure 2

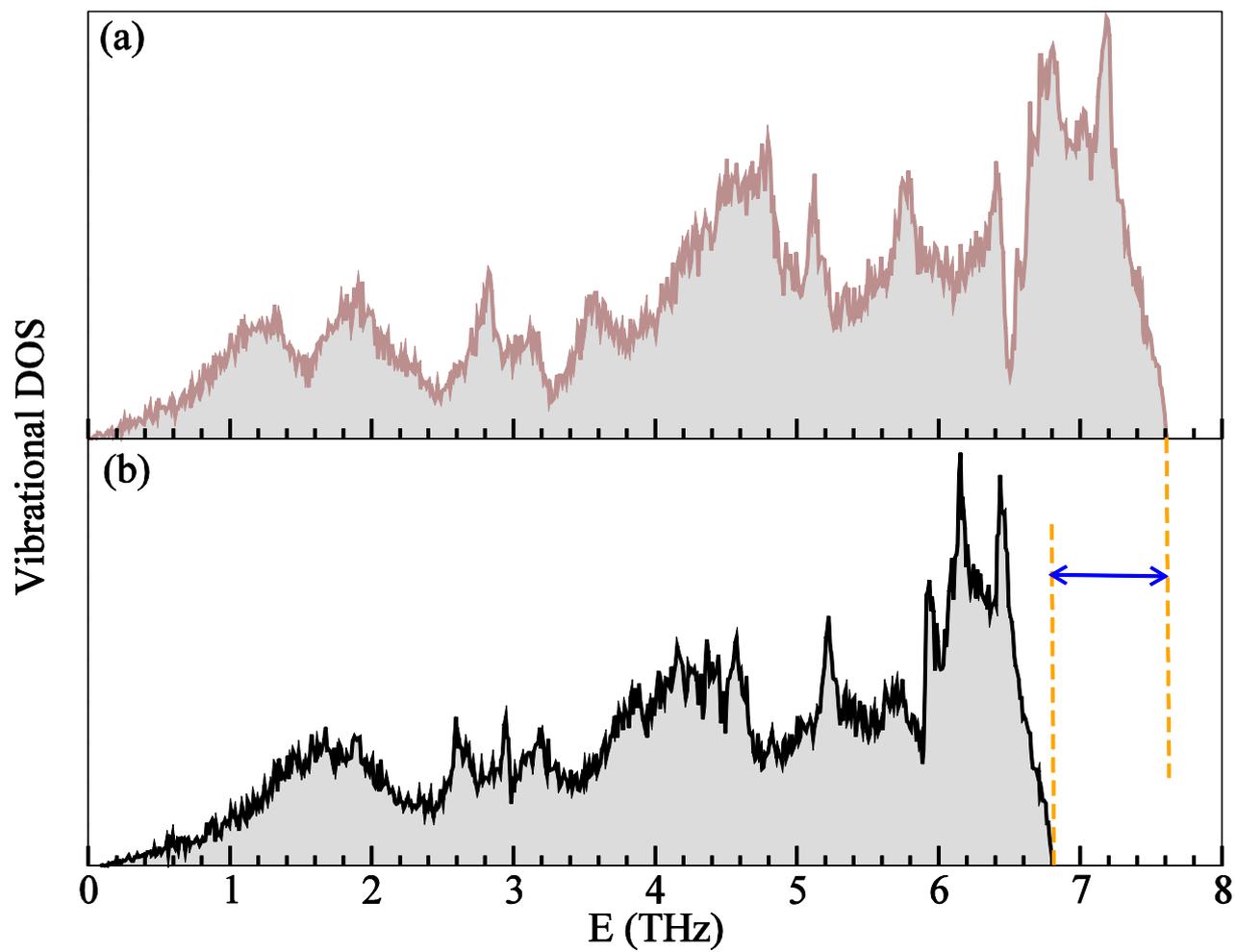

Figure 3

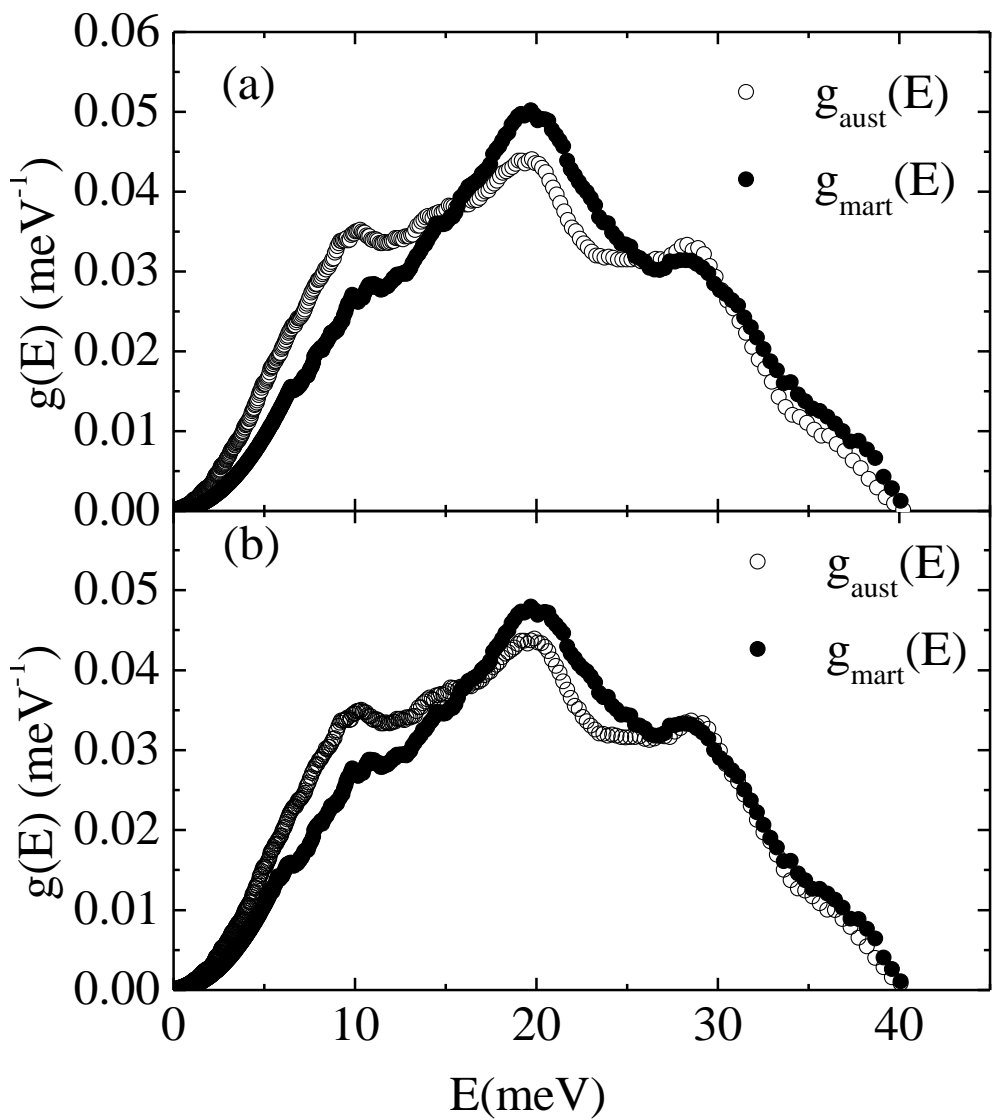

Figure 4